\documentstyle[11pt]{article}

\input{epsf.sty}

\begin{document}

\title{Magnetic Relaxation in the Peak Effect Region of CeRu$_2$\thanks{submitted to Phys. Rev. B}}

\author{Pei-Chun Ho\thanks{email:pch@physics.ucsd.edu},
        S. Moehlecke\thanks{Instituto de F$\acute{i}$sica ``Gleb
                            Wataghin'', Universidade Estadual de Campinas,
                            Unicamp, 13083-970, Campinas,
                            S$\widetilde{a}$o Paulo, Brasil.
                            email:sergio@ifi.unicamp.br},
        and M. B. Maple\thanks{email:mbmaple@ucsd.edu}
         \\Institute for Pure and Applied Physical Sciences
         \\and Department of Physics,
         \\University of California, San Diego
         \\La Jolla, CA 92093-0360, U.S.A.}

\date{\today}
\maketitle

\begin{abstract}

The different pinning strengths of the flux line lattice in the
peak effect (PE) region of a polycrystalline sample of CeRu$_2$
with a superconducting transition temperature \mbox{$T_c = 6.1$ K}
have been probed by means of magnetization measurements with a
SQUID magnetometer as the temperature $T$ and the magnetic field
$H$ are varied. Magnetic relaxation measurements were used to
monitor the flux line dynamics in the PE region. For \mbox{$T <
4.5$ K} and $H < H_P$, where $H_P$ is the field where the
magnetization reaches a maximum in the PE region, the relaxation
rate was found to be significantly larger in the descending-field
branch of the PE than it is in other sections of the PE region.
For \mbox{$T \geq 4.5$ K}, the relaxation rate in the entire PE
region is so large that the magnetization reached a stable
(equilibrium) value within \mbox{$10^4$ s}. This experimentally
determined stable state appears as an increase of the
magnetization in the PE region and has a dome shape superimposed
on a linear interpolation through the PE region. It was also found
that the PE in CeRu$_2$ can be suppressed by rapid thermal cycling
of the sample between \mbox{10 K} and \mbox{300 K} four times. The
reversible magnetization after the PE has been suppressed
coincides with the linear interpolation through the PE region, in
contrast to the behavior of the equilibrium magnetization when the
PE is present.


\end{abstract}


\section{Introduction}

The peak effect (PE) has been observed in elements such as
Nb,~\cite{Autler62} heavy fermion compounds such as
UPd$_2$Al$_3$,~\cite{Gloos93,Ishiguro95} UPt$_3$,~\cite{Tenya95}
the low superconducting critical temperature (T$_c$) C15 compound
CeRu$_2$,~\cite{Dilley96,Dilley97a} the layered compound
NbSe$_2$,~\cite{Bhattacharya93,Henderson96,Banerjee98,Paltiel00a,Paltiel00b,Paltiel00c}
the A15 compound V$_3$Si,~\cite{Isino88} and high T$_c$ materials
such as YBa$_2$Cu$_3$O$_{7-\delta}$,~\cite{Kwok94} and
Tl$_2$Ba$_2$CaCu$_2$O$_{8+\delta}$.~\cite{Hardy94} This phenomenon
occurs in type-II superconductors and is manifested as an increase
of the critical current density (i.e., the irreversibility of the
magnetization~\cite{Fietz69}) near the upper critical field
H$_{c2}$. The PE originates from the softening of the flux line
lattice (FLL) in the vicinity of H$_{c2}$ where the fluxoids are
more effectively pinned to randomly distributed pinning centers in
a sample, such as defects, impurity atoms, grain boundaries,
dislocations, etc.~\cite{Pippard69,Larkin79} However, the PE is
strongly material dependent, and the detailed understanding of
this phenomenon is still lacking.

The cubic Laves-phase (C15) compound CeRu$_2$ exhibits
superconductivity with a \mbox{T$_c \approx 6.1$ K}, the highest
value known for superconducting intermetallic compounds of Ce. It
was speculated that the PE in CeRu$_2$ is caused by a first-order
phase transition to the spatially nonuniform
Fulde-Ferrel-Larkin-Ovchinnikov (FFLO) superconducting
state,~\cite{Fulde64,Larkin65,Grunenberg96} which has also been
proposed to occur in the heavy fermion superconductors
UPd$_2$Al$_3$,~\cite{Gloos93,Steglich94}
UBe$_{13}$,~\cite{Thomas96} and UPt$_3$.~\cite{Tenya95} Extensive
transport,~\cite{Dilley96,Sato96}
magnetostriction,~\cite{Modler96} dc
magnetization,~\cite{Crabtree96,Roy98,Tulapurkar01,Banerjee98} ac
magnetic susceptibility,~\cite{Nakama95,Banerjee98} and neutron
scattering~\cite{Huxley96} measurements have been performed in the
PE region of CeRu$_2$. However, recent measurements of the
mixed-state flux-flow resistivity~\cite{Dilley96,Sato96,
Dilley97b} indicate that plastic deformation of the FLL may be
responsible for CeRu$_2$'s PE. In order to obtain more information
about the flux line dynamics associated with the PE, we have
performed measurements of the relaxation of the dc magnetization
$M$ as a function of temperature $T$, magnetic field $H$ and time
$t$ in PE region for CeRu$_2$.

\section{Experimental Details}

The as-cast large-grain polycrystalline CeRu$_2$ sample was
produced by a Czochralski pulling method.~\cite{Dilley96} The
sample has an irregular shape with dimensions $\sim$ \mbox{2 mm}
$\times$ \mbox{1.5 mm} $\times$ \mbox{1.5 mm}. An x-ray powder
diffraction analysis confirmed that the CeRu$_2$ sample has the
expected cubic C15 structure, although extra peaks due to Ru
inclusions were also present.

The dc magnetization data were obtained with a MPMS-5.5 (Quantum
Design, Inc.) superconducting quantum interference device (SQUID)
magnetometer. After the sample was cooled in zero field (ZFC), the
$M(T)$ data yielded \mbox{$T_c = 6.1$ K} with \mbox{$H = 100$ Oe}.
Special care was taken in the $M(T,H,t)$ measurements in the PE
region since it is known that the results of the measurements can
be affected by the measuring process~\cite{Ravikumar97} that
involves the movement of the sample in a magnetic field with small
inhomogeneity. The magnetization measurements for several
different scan lengths were tested, and a scan length of \mbox{1.5
cm} was found to minimize the effect of the inhomogeneous field of
the superconducting magnet without compromising significantly the
signal sensitivity. The field inhomogeneity for such a scan length
is \mbox{$\sim 0.004\%$}; i.e., \mbox{$\sim 1$ Oe} for a field of
\mbox{2 T}. Each measurement consisted of an average over two
scans (\mbox{$\sim$ 10 s/scan}) in the fixed-range mode which
prevented the performance of more than 2 scans in the
measurements. The sample was observed to be paramagnetic in the
normal state. Earlier resonant photoemission~\cite{Allen82} and
bremsstrahlung isochromat spectroscopy~\cite{Allen83} studies of
CeRu$_2$ have revealed a large amount of Ce 4f spectral weight in
the vicinity of the Fermi level. Both the Ce 4f electrons and the
excess Ru can contribute to the paramagnetic behavior of the
CeRu$_2$ sample. The paramagnetic background of our sample was
determined by measuring the magnetization $M(H)$ above $H_{c2}$
for each temperature. $H_{c2}$ is defined as the field where
$M(H)$ starts to deviate from the linear paramagnetic behavior in
the normal state. For the magnetization $M$ data presented here,
the linear paramagnetic background was removed and the data were
normalized to the sample weight (\mbox{47.62 mg}).

The magnetization hysteresis loop in the PE region is composed of
two branches: (1) the ascending-field branch where the flux lines
are moving into the sample with increasing field, and (2) the
descending-field branch where the flux lines are moving out of the
sample with decreasing field. For a given temperature and field in
the PE region, the relaxation of the magnetization was measured
for both ascending- and descending-field branches. The
measurements were performed as follows: In the ascending-field
branch, the sample was ZFC from the normal state to the measuring
temperature, $T_{M}$. The field was then raised in steps until it
reached the measuring field, $H_{M}$. While $H_{M}$ and $T_{M}$
were held constant, the magnetization was measured as a function
of time. For the descending-field branch, the field was either
increased to \mbox{$\sim 2$ T} above H$_{c2}$ at $T_{M}$ or the
sample was FC (field cooled) from 10K to $T_{M}$ at $H \geq
H_{c2}(T_{M})$ to ensure the full flux penetration (i.e., the
descending-field branch always started from the normal state of
CeRu$_2$). The field was then lowered by steps to $H_{M}$ to
measure the relaxation of the magnetization. Because the PE region
is strongly history-dependent, the field steps for increasing and
decreasing the field were kept the same to maintain the same
initial conditions for each relaxation measurement. Between each
pair of relaxation measurements (ascending and descending branches
with the same $T$), the temperature of the sample was raised to
\mbox{10 K} ($> T_c$) and the remanent field of the magnet was
minimized by oscillating the field from \mbox{2 T} to \mbox{0 T}.

\section{Results}

\begin{figure}
 \begin{center}
  \epsfbox{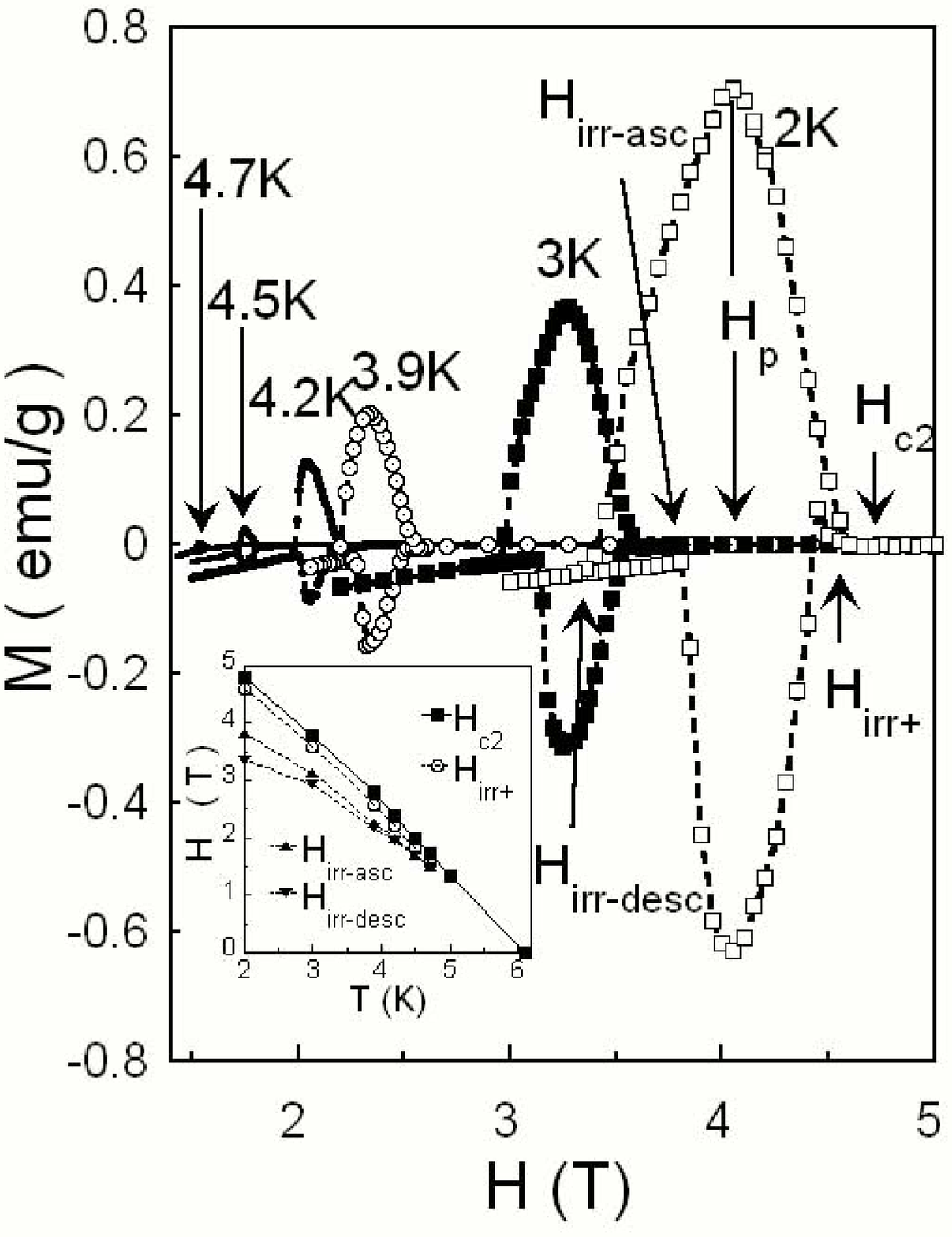}
 \end{center}
 \caption{A set of hysteresis loops in dc magnetization $M(H)$ of
          polycrystalline CeRu$_2$ in the PE region between
          \mbox{2 K} and \mbox{4.7 K}. Inset: $H-T$ diagram of
          CeRu$_2$. $H_{irr+}$ is defined as the field where PE
          disappears, $H_{irr-asc}$ as the onset field in the
          ascending-field branch, $H_{irr-desc}$ as the onset field
          in the descending-field branch, and $H_p$ as the field where
          the maximum of the irreversible magnetization occurs. The
          mismatch between $H_{irr-asc}$ and $H_{irr-desc}$ becomes
          more pronounced with decreasing temperatures.}
 \label{fig:PEloops}
\end{figure}

The magnetic field dependencies of the magnetization $M(H)$ in
CeRu$_2$'s PE region for various temperatures between \mbox{2 K}
and \mbox{5 K} are shown in Fig.~\ref{fig:PEloops}. The irreversible
peak feature in $M(H)$ appears between $H_{irr-}$ and $H_{irr+}$,
the low and high fields where magnetization hysteresis occurs.
However, when \mbox{$T <$ 4.2 K}, $H_{irr-}$ appears in different
fields for the ascending- and descending-field branches (defined as
$H_{irr-asc}$ and $H_{irr-asc}$, respectively). With increasing
magnetic field, the PE starts at $H_{irr-asc}$, whereas with
decreasing field from above $H_{c2}$, the magnetization hysteresis
persists down to $H_{irr-desc} < H_{irr-asc}$. This difference in
$H_{irr-asc}$ and $H_{irr-desc}$ is not observed in the magnetization
data $M(H)$ above \mbox{ 4.5 K} due to the fast relaxation
rates at these temperatures as we will show later. As for $H_{irr+}$,
there is no difference in this quantity between the two branches.
The location of the PE region and $H_{c2}$ are displayed in the $H-T$
phase diagram in the inset of Fig.~\ref{fig:PEloops}. As $T$ increases,
the height and width of the hysteresis in $M(H)$ gradually
decrease. The PE becomes undetectable within the resolution of the MPMS
for \mbox{$T > 4.7$ K}. For each PE hysteresis loop, the beginning of the
irreversibility corresponds to an increase in the critical current
density \mbox{$J_c \propto \Delta M (\equiv M_{desc}(H) - M_{asc}(H)$}
where $M_{asc/desc}$ is the magnetization from the ascending/descending-field
branch) which is a result of the increase of the effective pinning strength
of the FLL. After the field passes through $H_P$ (the field where the
maximum magnetization hysteresis occurs within the PE region), the
subsequent reduction in size of the hysteresis loop of the PE corresponds
to a decrease of the FLL pinning, yielding a reduction in $J_c$.

\begin{figure}
 \begin{center}
  \epsfbox{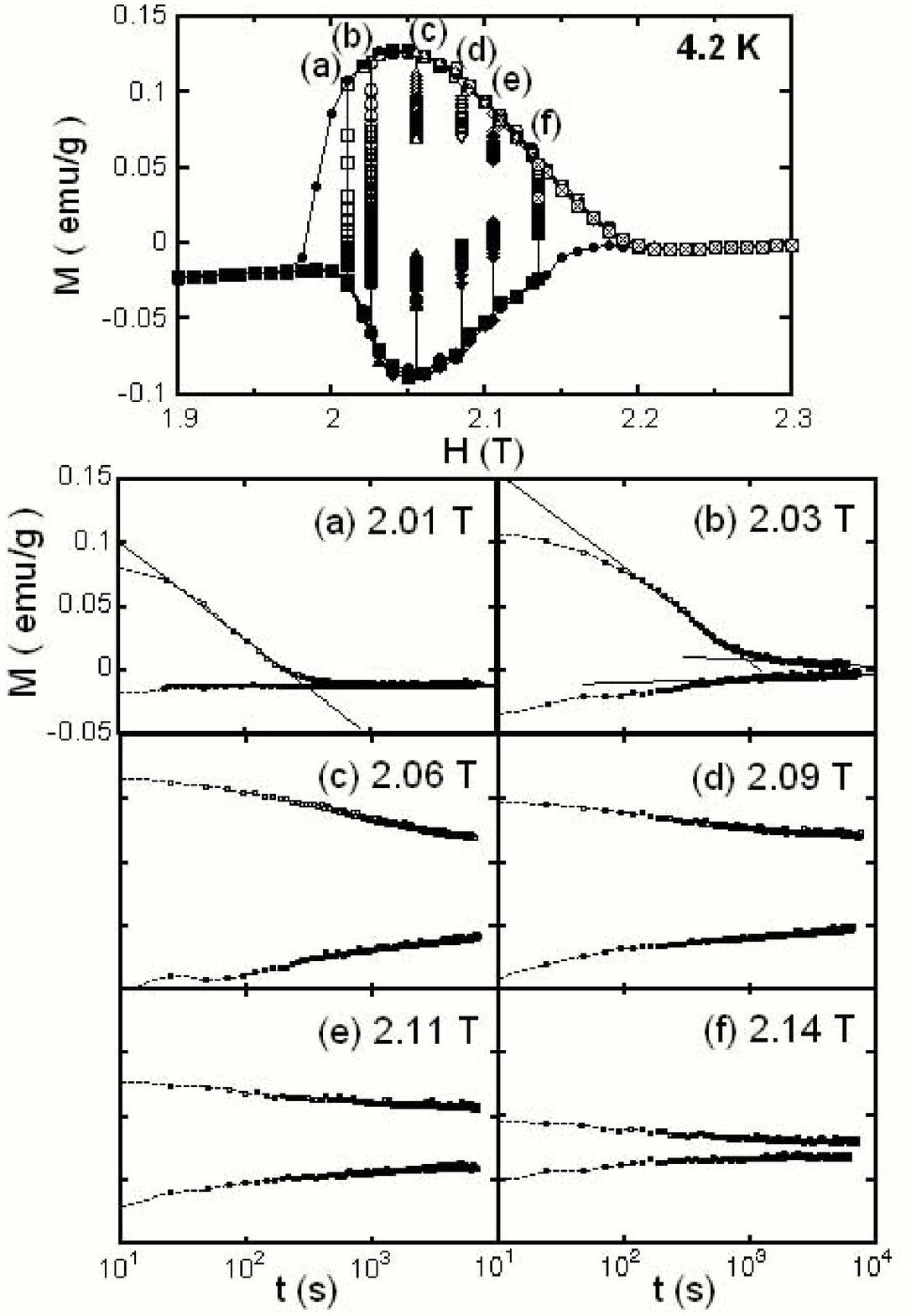}
 \end{center}
 \caption{Relaxation of the magnetization $M$ of CeRu$_2$ at six different
          fields $H$ in the PE region at \mbox{4.2 K}.
          $M$ is plotted vs H in the top panel.
          Panels (a)-(f) show $M$ vs $t$ at
          each $H$ and all have the same vertical and horizontal scales.
          The bottom curve is from the ascending-field branch,
          and the top curve is from the descending-field branch.
          The slopes ($dM(emu/g)/d(lnt)$) of the straight
          lines are: panel(a), \mbox{$\sim -3.34\times 10^{-2}$}
          and \mbox{$\sim 2.15\times 10^{-4}$}; panel(b),
          \mbox{$\sim -3.32\times 10^{-2}$},
          \mbox{$\sim -2.16\times 10^{-3}$},
          and \mbox{$\sim 1.46\times 10^{-3}$}.}
 \label{fig:rlx4p2K}
\end{figure}

\begin{figure}
 \begin{center}
  \epsfbox{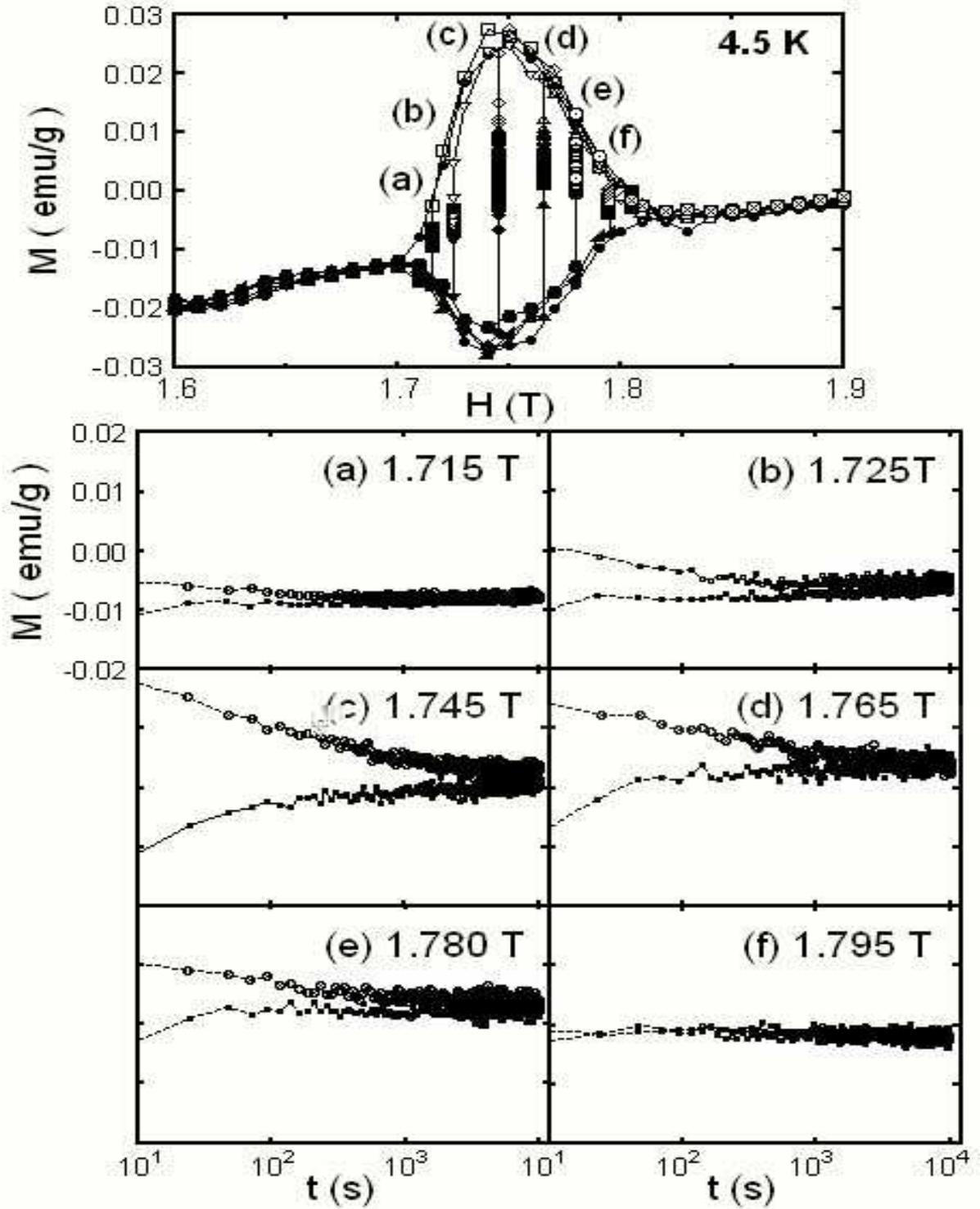}
 \end{center}
 \caption{Relaxation of the magnetization $M$ of CeRu$_2$ at six different magnetic fields
          in the PE region at \mbox{4.5 K}. The $M$ is plotted vs $H$ in
          the top panel.
          Panels (a)-(f) show $M$ vs $t$ at each $H$ and all have the
          same vertical and horizontal scales.
          The bottom curve is from the ascending-field branch,
          and the top curve is from the descending-field branch.}
 \label{fig:rlx4p5K}
\end{figure}

The measurements of the isothermal relaxation of the magnetization of
CeRu$_2$ were performed at constant fields within the PE region for
T between \mbox{2 K} and \mbox{5 K}. Shown in Figs.~\ref{fig:rlx4p2K}
and ~\ref{fig:rlx4p5K} are the raw magnetization relaxation data
$M(H,t)$ for six different fields at \mbox{4.2 K} and \mbox{4.5 K},
respectively. The top panels in Figs.~\ref{fig:rlx4p2K} and
~\ref{fig:rlx4p5K} show the evolution of the magnetization with
time in the $M-H$ plane, and panels (a)-(f) depict the time dependence
of the magnetization $M(t)$ after a specified $H$ was reached in both
branches at \mbox{4.2 K} and \mbox{4.5 K}. At \mbox{4.2 K} and
\mbox{2.01 T}, the relaxation rate ($\arrowvert dM/d(lnt)\arrowvert$)
in the descending-field branch changes from
\mbox{$\sim 3.34\times 10^{-2}$} to \mbox{$\sim 2.32\times 10^{-4}$}
after \mbox{300 s}; the ascending- and descending-field magnetization
after \mbox{300 s} merges to the same value (Fig.~\ref{fig:rlx4p2K}(a)).
At a different $H$ of \mbox{2.025 T} (Fig.~\ref{fig:rlx4p2K}(b))
(closer to \mbox{$H_P$ = 2.05 T}), the descending-field relaxation rate
changes from \mbox{$\sim 3.23\times 10^{-2}$} to
\mbox{$\sim 2.16\times 10^{-3}$} after \mbox{1000 s} and the
ascending-field rate after \mbox{500 s} is
\mbox{$\sim 1.46\times 10^{-3}$}. The magnetizations in each branch
nearly merged to a stable value within \mbox{$10^4$ s}. For
$H \geq H_P$, the magnetization in each branch depends approximately
linearly on the logarithm of time. However, the magnetization in these
fields cannot reach a stable value within \mbox{$\sim 10^4$ s}.
From the intersection of two straight lines that were fitted to
the $M(logt)$ data, the time of \mbox{$\sim 2\times 10^6$ s} was
estimated for the magnetization to reach a stable value in the PE region
$H \geq H_P$ at \mbox{4.2 K}. Thus, at \mbox{4.2 K} in
the PE region, for $H < H_P$, the relaxation in the descending-field
branch is dramatically faster than it is in the ascending-field branch.
Compared to \mbox{4.2 K} data, the relaxation at \mbox{4.5 K} is even
faster. The \mbox{4.5 K} magnetizations from each branch can merge to a
stable value within \mbox{3000 s} across the entire PE region
(Fig.~\ref{fig:rlx4p5K}).

\begin{figure}
 \begin{center}
  \epsfbox{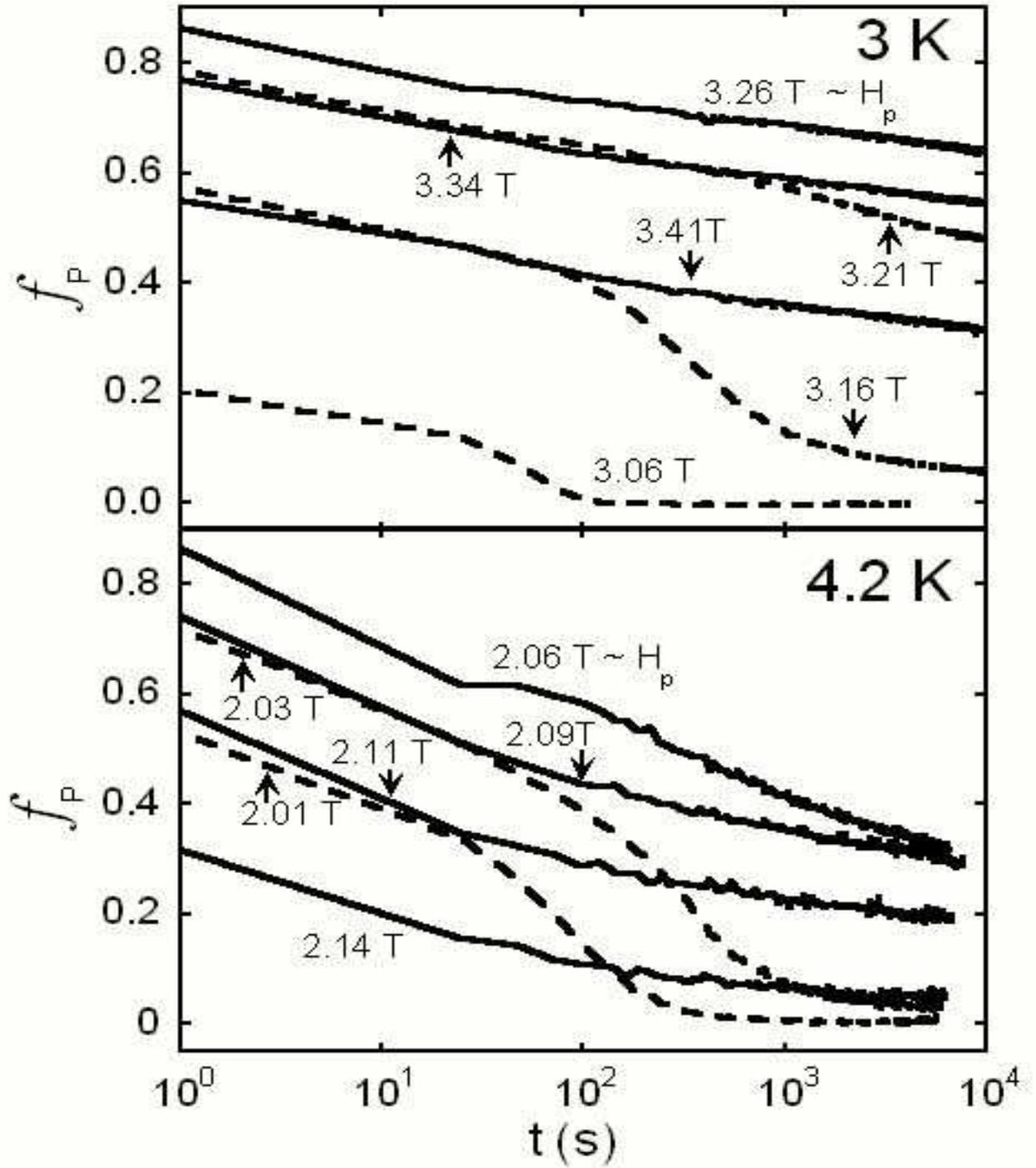}
 \end{center}
 \caption{Time dependence of the normalized pinning force $f_p$
          at six different fields for \mbox{3 K} and \mbox{4.2 K},
          respectively.}
 \label{fig:fplnt}
\end{figure}

Because \mbox{$J_c \propto \Delta M$}, the flux pinning force
density \mbox{$F_p = J_c \times B/c$} and $B \sim H$, it follows
that$F_p \propto$ \mbox{$\Delta M \times H$}. Fig.~\ref{fig:fplnt}
illustrates the normalized effective pinning force $f_p$ as a
function of time at \mbox{3 K} in (a) and \mbox{4.2 K} in (b)
where $f_p$ is defined as \mbox{$(\Delta M/\Delta M_{max-init})
\times (H/H_{c2})$} and $\Delta M_{max-init}$ is the initial size
of $\Delta M$ at $H_P$. As the magnetic field approaches $H_P$,
$f_p$ becomes stronger and reaches a maximum at $H_P$. Although
the initial value of $f_p$ is the same for two different values of
H above and below $H_P$, $f_p$ for $H < H_P$ relaxes much faster
than $f_p$ for $H > H_P$. Hence, $H \geq H_P$ is a stronger
pinning region of the PE, and $f_p$ has more linear logarithmic
time dependence here. The $H \geq H_P$ data at \mbox{3 K} can be
well represented by expression for $f_p$ from the Anderson-Kim
thermally activated flux creep model, \mbox{$f_p
\propto[1-(k_BT/U_0)lnt]$} where $k_B$ is Boltzmann's
constant.~\cite{Anderson62,Kim62,Anderson64,Tinkham2ed} The
activation energy $U_0$ for fluxoid depinning is estimated to be
\mbox{$\sim 115$ K}. For \mbox{$T=4.2$ K} and  $H > H_P$, the
relaxation deviates from the linear logarithmic time dependence
and $f_p$ can be described with the collective flux creep
model~\cite{Feigel'man89} which yields \mbox{$f_p \propto
1/[1-(\mu k_BT/U_0)ln(t/t_0)]^{(1/\mu)}$} where $t_0$ is a
macroscopic quantity depending on the sample size and $\mu$ comes
from $U(J) \propto J^{-\mu}$. For $H =$ 2.085 T, 2.105 T, and
2.135 T, $\mu \sim$ 0.48, 0.73, and 0.78, respectively, which are
reasonable values as expected from the collective creep model. The
relaxation data for $H < H_P$ at both \mbox{$T=3$ K} and \mbox{4
K}, the relaxation data cannot be described by any known model.

\begin{figure}
 \begin{center}
  \epsfbox{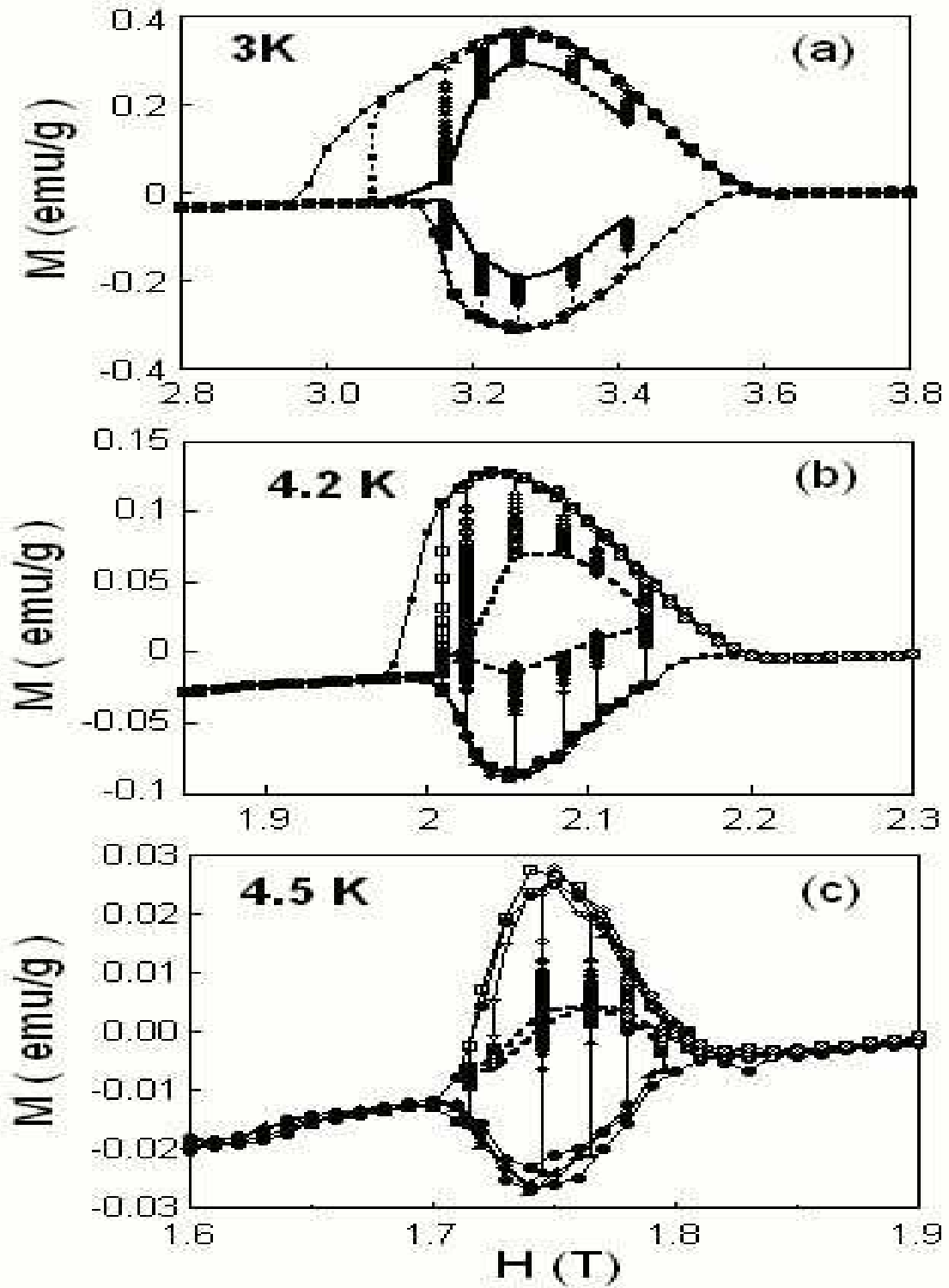}
 \end{center}
 \caption{Evolution of the PE hysteresis loops in $M$ vs $H$
          at \mbox{3 K}, \mbox{4.2 K}, and \mbox{4.5 K}
          (\mbox{$\sim 24$ s} between the data points at each
          $H$).
          The dotted lines are the connection between the data
          points at \mbox{$\sim 10^4$ s} after the initial state.
          Note the disappearance of the hysteresis loop at
          \mbox{4.5 K}.}
 \label{fig:minorPEs}
\end{figure}

Fig.~\ref{fig:minorPEs} shows the time evolution of magnetization
hysteresis loops $M(H)$ of the PE at \mbox{3 K}, \mbox{4.2 K}, and
\mbox{4.5 K}. The dotted line connects the final values of the
magnetization at the end of the relaxation measurements at each
temperature (i.e., \mbox{$\sim 10^4$ s} after the initial state).
For \mbox{$T \leq 4.2$ K}, the hysteresis loops of the PE evolve
toward smaller loops, but for \mbox{$T \geq 4.5$ K}, the
hysteresis loops disappear after \mbox{$\sim 3000$ s} and the
magnetization from both branches reaches a stable value within the
PE region.

\begin{figure}
 \begin{center}
  \epsfbox{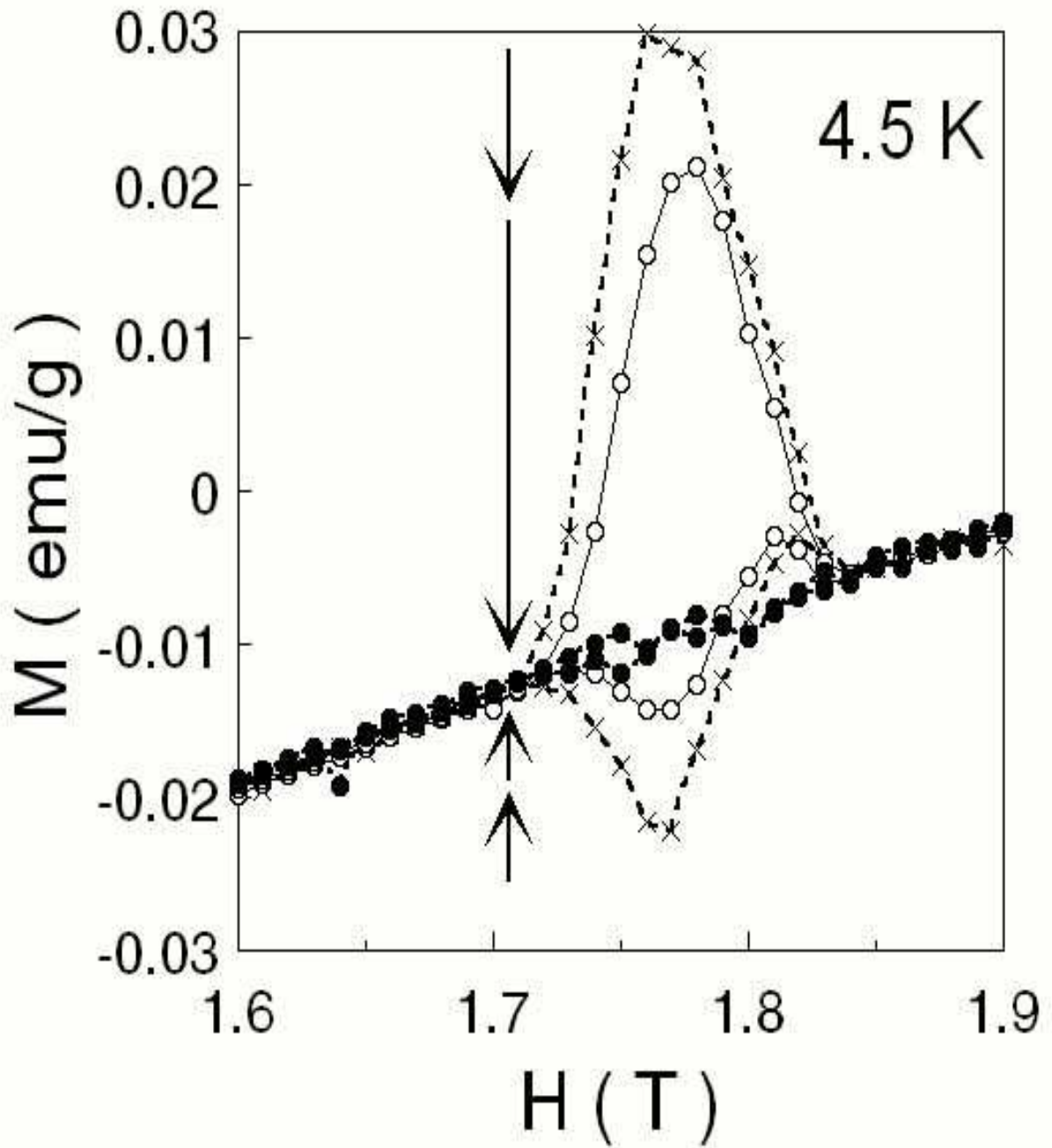}
 \end{center}
 \caption{Gradual destruction of the PE hysteresis loops $M(H)$
          at 4.5 K after several rapid thermal cycles between
          \mbox{300 K} and \mbox{4.5 K}. The magnetization
          ($\bullet$) became reversible and different from the
          stable state magnetization observed in Fig.6(a).}
 \label{fig:dstroyPE}
\end{figure}

Since for \mbox{$T \geq 4.5$ K} the magnetization from both
branches reaches about the same value after \mbox{$\sim 3000$ s},
a stable value of the magnetization of CeRu$_2$ at each field can
be determined by averaging the last \mbox{300 s} of magnetization
data $M(t)$ at \mbox{4.5 and 4.7 K}. Fig.~\ref{fig:eqPEs} depicts
the experimentally determined stable state of the magnetization of
CeRu$_2$ ($M_{stable}$) within the PE. This experimentally determined
$M_{stable}$ is dome-shaped with a maximum slightly above $H_P$.
However, we also found the PE of CeRu$_2$ at 4.5 K can be suppressed
by rapid thermal cycling of the sample between \mbox{10 K} and
\mbox{300 K} several times. After the PE is destroyed at \mbox{4.5 K},
the magnetization becomes reversible and follows approximately the
linear interpolation between the beginning and the end of the initial
hysteresis loop of the PE (Fig.~\ref{fig:dstroyPE}).

\section{Discussion}


Previous measurements~\cite{Dilley96,Sato96,Dilley97b} of the flux
flow resistance $R(H)$ in the mixed state of CeRu$_2$ showed field
hysteresis at the lower onset of the PE region: for $J < J_c$,
near the onset of the PE the resistance is higher in the ascending
field branch than in the descending-field branch. The flux motion
is inhibited more in the descending-field branch near the lower
onset of the PE. This behavior is equivalent to the observed
mismatch $H_{irr-desc} < H_{irr-asc}$ in the $M(H)$ data for
\mbox{$T \leq 4.2$ K} (Fig.~\ref{fig:PEloops}), except at the same
temperature it is more pronounced in resistivity data than the
magnetization data.~\cite{Dilley96} This hysteresis at the lower
onset of the PE was also observed in several superconducting
materials, such as 2H-NbSe$_2$,~\cite{Banerjee98,Ravikumar00b}
Ca$_3$Rh$_4$Sn$_{13}$,~\cite{Sarkar00} and Nb.~\cite{Kopelevich98}
Also shown by the relaxation data of the magnetization
(Fig.~\ref{fig:rlx4p2K}), the flow of the FLL in this portion of
the PE changes very quickly from the behavior of a softer FLL
(i.e., larger pinning) in the beginning to more coherent motion of
a stiff FLL (i.e., reduction of pinning). Thus, this hysteresis
seems strongly connected to the metastable disordered phase
generated by edge contamination observed in
2H-NbSe$_2$.~\cite{Paltiel00a,Paltiel00b,Paltiel00c}. Furthermore,
it is believed that bulk pinning is greatly reduced in more
2-dimensional compound 2H-NbSe$_2$.~\cite{Paltiel00c} However, the
similarity of behavior of the fluxoids in the PE region of the
cubic CeRu$_2$ compound to that of NbSe$_2$ indicates that the
underlying pinning mechanism may be independent of crystal
dimensionality.

\begin{figure}
 \begin{center}
  \epsfbox{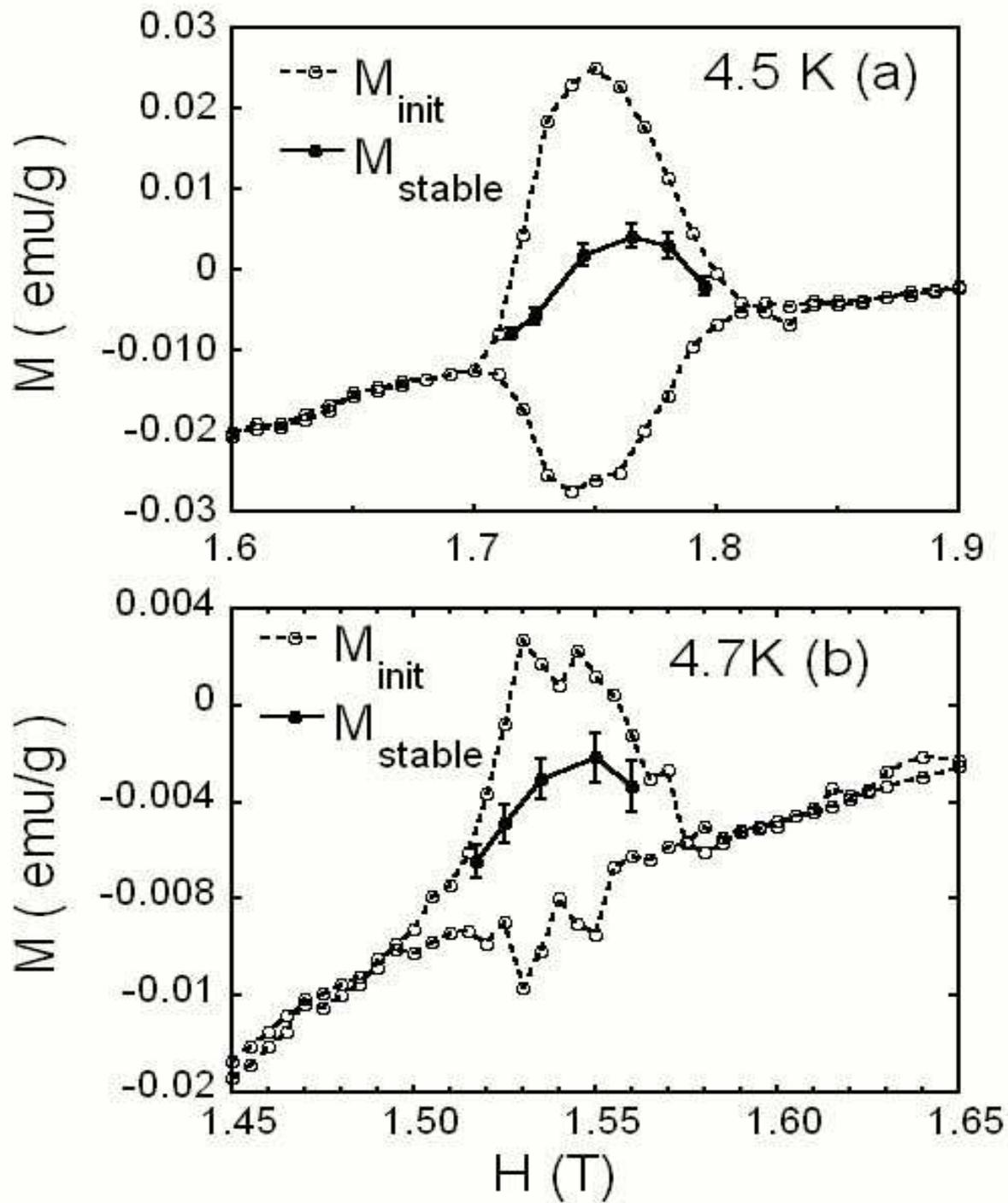}
 \end{center}
 \caption{Initial hysteresis loop $M_{init}(H)$ of the PE and the
          stable state magnetization ($M_{stable}(H)$) determined
          from the relaxation data at \mbox{4.5 K}(a) and
          \mbox{4.7 K}(b).}
 \label{fig:eqPEs}
\end{figure}

As observed from the relaxation data, the FLL in the disordered
state for \mbox{$T < 4.5$ K} and $H > H_P$ has not yet reached a
thermal equilibrium state since its effective pinning force is
still decreasing towards zero. As \mbox{$T \geq $ 4.5 K}, vortex
lines can reach an equilibrium state by fast relaxation that is
possibly due to thermal fluctuations or to the shaking
effect~\cite{Giamarchi96,Paltiel00b}caused by the scanning
movement of the sample from one pick up coil to the other in SQUID
magnetometer. The stable-state magnetization
(Fig.~\ref{fig:eqPEs}) we determined from magnetic relaxation in
the PE of CeRu$_2$ is reminiscent of a phase transition,
especially compared with the magnetization data taken in the same
sample after its PE is suppressed (Fig.~\ref{fig:dstroyPE}).
Recently, two experiments were reported in which the stable-state
magnetization of CeRu$_2$ in the PE region was determined by
cycling the magnetic field with a small amplitude at \mbox{4.5 K}
in CeRu$_2$'s PE.~\cite{Roy98,Tulapurkar01} The stable-state
magnetization derived from our relaxation measurements is
straightforward and does not depend by any
model,~\cite{Ravikumar00a} in contrast to the
magnetic-field-cycling experiments. The stable-state magnetization
determined by Roy et al.~\cite{Roy98} has the same dome shape as
found in our equilibrium magnetization, but the magnetization
found by Tulapurkar et al.~\cite{Tulapurkar01} has a step shape.
Although these experimental results do not agree with one another,
in both cases this stable state was considered to result from an
occurrence of a first order phase transition.

If the stable-state magnetization at \mbox{4.5 K} and \mbox{4.7 K}
is assumed to be related to the first order melting of the FLL,
then the jump in magnetization expected for such a vortex melting
transition can be estimated from the Clausius-Claperon relation,
$\Delta S = -\Delta M_{melting} dH_m / dT$.~\cite{Schilling96a}
The entropy change at the transition per unit volume is $\Delta S
\approx c_L^2C_{66} / T_m$,~\cite{Zeldov95} where $c_L \approx
0.2$ is the Lindemann number and $C_{66} \approx
[BH_{c1}/(16\pi)](1-b)^2$ is the vortex-lattice shear modulus for
an isotropic superconductor~\cite{Tinkham2ed} where \mbox{$b
\equiv B/H_{c2}$}.
At \mbox{$T_m = 4.5$ K}, \mbox{$B \sim H_P = 17.55$ kOe},
\mbox{$H_{c1} \sim 300$ Oe} and \mbox{$H_{c2} \sim 19.8$ kOe},
this gives \mbox{$C_{66} \approx 1353$ erg/cm$^3$} and
\mbox{$\Delta S \approx 12.03$ erg/K-cm$^3$}. From $M_{stable}(H)$
data at \mbox{4.5 K} and \mbox{4.7 K} (Fig.~\ref{fig:eqPEs}),
$dH_m/dT$ is assumed to be the same as \mbox{$dH_P/dT \approx
-10.75$ kOe/K} and $\Delta M_{melting}$ is estimated to be
\mbox{$1.12 \times 10^{-3}$ emu/cm$^3$}; i.e., \mbox{$1.05 \times
10^{-4}$ emu/g} where the density~\cite{Modler96} of CeRu$_2$ is
\mbox{10.62 g/cm$^3$}. This estimate of $\Delta M_{melting}$ is
much smaller than what we observed from \mbox{$\Delta M$ $\sim 1
\times 10^{-2}$ emu/g} at \mbox{4.5 K} and \mbox{$\sim 4 \times
10^{-3}$ emu/g} at \mbox{4.7 K} (Fig.~\ref{fig:eqPEs}). Whether
the stable-state magnetization is associated with a first order
melting in the the PE region but is obscured by the disordered
phase appearing near the sample surface, the magnetization data
alone are not sufficient for resolving this issue. Thus, whether
the FLL undergoes a first-order or continuous phase transition in
the PE region or simply just a crossover of the dynamical behavior
is not yet clear.


How CeRu$_2$'s PE can be suppressed by rapid thermal cycling of
the temperature between 10 K and \mbox{300 K} is still a mystery.
Previous low temperature X-ray diffraction studies show the Laves
phase compounds LaRu$_2$~\cite{Lawson74} and
(La$_{1-x}$,Ce$_x$)Ru$_2$~\cite{Shelton77} with $x < 0.25 $
undergo a cubic-tetragonal structural transition at \mbox{$\sim
30$ K}. We suspected a small part of the CeRu$_2$ sample might
also undergo a similar structural phase transition at some
temperature between 10 K and 300 K. Small regions of transformed
structure within the sample could provide pinning centers. If the
sample temperature was changed rapidly, the high temperature Laves
phase crystal structure of CeRu$_2$ might be quenched to low
temperature and reduce the PE. However, the previous X-ray studies
did not provide any evidence of such a structural phase
transformation in CeRu$_2$~\cite{Shelton77}. After suppressing the
PE, we also thermally cycled the sample at a rate of \mbox{10
K/min} to test whether some significant change in magnetization in
the PE region at certain temperature between \mbox{10 K} and
\mbox{300K} would occur in support of our hypothesis. However,
this experiment did not yield conclusive results.

\section{Summary}
In summary, we have performed magnetic relaxation measurements
between \mbox{2 K} and \mbox{4.7 K} in the PE region of CeRu$_2$.
For \mbox{$T \leq 4.2$ K}, due to stronger pinning, the FLL
relaxes at a much slower rate and does not reach a stable state
within the duration of the measurement (\mbox{$\sim 1.2\times
10^4$ s}). The dynamics of the FLL for $H < H_P$ in the
descending-field branch are very different than in the other parts
of the PE region, and the pinning strength is stronger for
\mbox{$H > H_P$} than for \mbox{$H < H_P$}. For \mbox{$T\geq 4.5$
K}, a stable-state magnetization exists across the PE region that
has a dome shape with a maximum occurring slightly above $H_P$ and
agrees with the result of magnetic-field-cycling measurements by
Roy et al.~\cite{Roy98} We also found that the PE of CeRu$_2$ can
be removed by rapid thermal cycling of the sample between room
temperature and \mbox{10 K} several times, and the reversible
magnetization follows a linear interpolation across the PE region.
Compared with the reversible magnetization of CeRu$_2$ without the
PE, the stable-state magnetization associated with the appearance
of the PE is reminiscent of a phase transition. However, the size
of the stable-state magnetization jump at $H_P$ is much larger
than expected for a first order melting of the FLL. Whether this
stable-state magnetization is related to a first-order or a
continuous phase transition, or simply a dynamical crossover of
the FLL is not yet clear.


\section{Acknowledgement}
We thank Professor Daniel Arovas for helpful discussions. This research
was supported by the \mbox{U. S.} Department of Energy under Grant No.
DEFG-03-86ER-45230.


\newpage

\epsfxsize=500pt \epsfysize=600pt

\epsfxsize=500pt \epsfysize=600pt

\epsfxsize=500pt \epsfysize=600pt

\epsfxsize=500pt \epsfysize=600pt

\epsfxsize=500pt \epsfysize=600pt

\epsfxsize=500pt \epsfysize=600pt

\epsfxsize=500pt \epsfysize=600pt








\end{document}